\documentclass[reprint,superscriptaddress,amsmath,amssymb,amsart,aps,prb]{revtex4-2}
\usepackage{accents}
\usepackage{graphicx}
\usepackage{dcolumn}
\usepackage{bm}
\usepackage{braket}
\usepackage{amsthm}
\usepackage{bookmark} 
\usepackage[autostyle]{csquotes}
\usepackage{hyperref}
\usepackage{booktabs}
\graphicspath{{../figs/}}
\usepackage{lineno}

\begin{document}

\title{Spin textures at the Fermi surface of ferromagnets and associated torque}

\author{Thierry Valet}
\email[Corresponding author :]{ tvalet@mphysx.com}
\affiliation{MPhysX O\"U, 15551 Tallinn, Estonia}
\affiliation{Advanced Institute of Materials Research (WPI-AIMR), Tohoku University, Sendai 980-8577 Japan}
\affiliation{Center for Science and Innovation in Spintronics (CSIS), Tohoku University, Sendai 980-8577 Japan
}

\author{Jing Li}
\affiliation{Universit\'e Grenoble Alpes, CEA, LETI, 38054 Grenoble, France}

\author{Henri Jaffr\`es}
\affiliation{Laboratoire Albert Fert, CNRS, Thales, Universit\'e Paris-Saclay, 91767 Palaiseau, France}

\author{Vincent Cros}
\affiliation{Laboratoire Albert Fert, CNRS, Thales, Universit\'e Paris-Saclay, 91767 Palaiseau, France}

\author{Mairbek Chshiev}
\affiliation{Universit\'e Grenoble Alpes, CEA, CNRS, Spintec, 38054 Grenoble, France}

\author{Roberto Raimondi}
\affiliation{Dipartimento di Matematica e Fisica,  Universit\`a  Roma Tre, Via della Vasca Navale 84, 00146 Roma, Italy}

\date{\today}

\begin{abstract}
We establish theoretically that transverse spin textures are generically induced by the spin-orbit interaction at the Fermi surface of centrosymmetric ferromagnetic metals. We confirm this prediction with {\em ab initio} computations for Fe, Co and Ni. By symmetry, this cannot be associated with spin-orbit torques under a uniform current. However, a new {\em current-gradient-induced} torque emerges in nanoscale geometries. For a series of Ni-based bilayers, the computed Ni thickness dependence of the current-gradient-induced torque is in quantitative agreement with measurements previously attributed to orbital torques. We discuss how this opens a new avenue for the electrical control of magnetic orders, possibly also including the N\'eel vector of metallic altermagnets.
\end{abstract}

\maketitle

\setlength\linenumbersep{5pt}

In ferromagnetic metals, when neglecting the spin-orbit interaction (SOI), each band is either a majority or minority spin band. Hence, the ones crossing the Fermi level define Fermi surfaces of pure spin character. Local degeneracies between bands of opposite spin are generic and occur on two-dimensional surfaces in the Brillouin zone~\cite{neumann:1929,herring:1937,gosalbez:2015}, resulting in majority and minority Fermi surfaces intersecting along lines. It is known that the SOI lifts these degeneracies and hybridizes the spin states, leading to reconnected Fermi surfaces \cite{wang:2007,gradhand:2009}. We call them {\em spin mosaics}, their extended regions of nearly pure majority or minority states, {\em tiles}, and the narrow regions where the tiles meet, {\em seams}. The role of these SOI-induced hybridizations is well understood for the anomalous Hall effect \cite{yao:2004,wang:2006}, spin relaxation \cite{mokrousov:2013}, and interfacial magnetic anisotropy \cite{yang:2011,dieny:2017}. However, crucially, it was previously assumed that ``the spin polarization vanishes at the splitting points'' \cite{gradhand:2009} (see also Fig.~4b of Ref.~\cite{amin:2019}).  

\begin{figure}[!t]
\vspace*{2.6 pt}
\includegraphics[width=0.95\columnwidth]{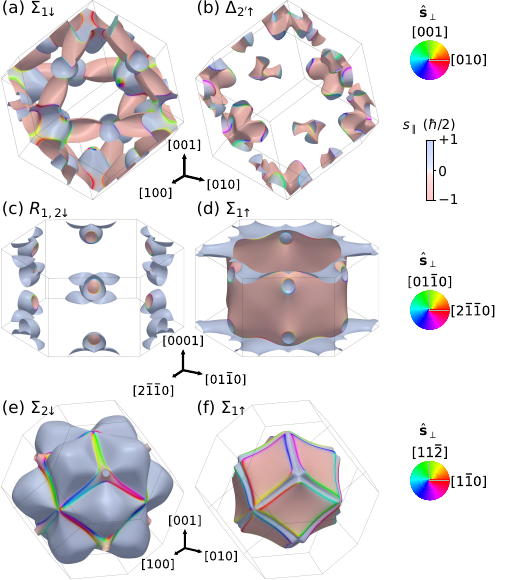}
\vspace{-7.6 pt}
\caption{\label{fig:1} {\em Spin mosaics of Fe, Co and Ni.} We show here pairs of Fermi surfaces, which intersect in the absence of SOI, and carry the strongest transverse spin texture: in bcc Fe with the magnetization direction ${\bm m}$ along $[100]$ {\em i.e.}, (a) $\Sigma_{1\downarrow}$ and (b) $\Delta_{2'\uparrow}$; in hcp Co with ${\bm m}$ along $[0001]$ {\em i.e.}, (c) $R_{1,2\downarrow}$ and (d) $\Sigma_{1\uparrow}$; in fcc Ni with ${\bm m}$ along $[111]$ {\em i.e.}, (e) $\Sigma_{2\downarrow}$ and (f) $\Sigma_{1\uparrow}$. Each surface is named after its dominant ``parent'' band before reconnection, in the notation of Ref.~\cite{bouckaert:1936} for Fe and Ni and of Ref.~\cite{bradley:1972} for Co, and is drawn inside the first Brillouin zone. The red tint, on the majority tiles, and blue tint, on the minority ones, give the spin polarization along ${\bm m}$. It fades to white at the seams, where the hue from the color wheels gives the direction of the transverse spin polarization.}
\end{figure}

In this Letter, we demonstrate that in fact a spin texture, transverse to the magnetization, generically exists at the seams in centrosymmetric ferromagnets. We confirm this with {\em ab initio} calculations for the elemental ferromagnets bcc Fe, hcp Co, and fcc Ni, as shown in Fig.~\ref{fig:1}. Spin textures in momentum space have been previously discussed in relation to systems with broken inversion symmetry, where they are associated with spin-orbit torques (SOTs) under a uniform current \cite{edelstein:1990,manchon:2008,chernyshov:2009,garate:2009,miron:2010}, forbidden here. Instead, we establish that a new current-gradient-induced SOT appears, accounting quantitatively for measurements on Ni-based samples~\cite{hayashi:2023}, which have been interpreted as evidence of a torque induced by the orbital Hall effect (OHE)  \cite{bernevig:2005,kontani:2009,go:2018,go:2020}. We also discuss broader implications in relation to spin-vorticity coupling (SVC) \cite{matsuo:2011,okano:2019,nakayama:2025,yi:2025}, to altermagnets \cite{pekar:1965,noda:2016,hayami:2019,yuan:2020,smejkal:2022,yang:2025}, and to the future of non-volatile electronics \cite{dieny:2020,fert:2024}.

{\em Transverse spin texture.---} We consider a centrosymmetric ferromagnetic metal for which, in the absence of SOI, the Fermi surfaces of a certain majority band and a certain minority band intersect. We introduce the corresponding eigenstates as  \smash{$|a\!\uparrow\rangle = |a\rangle \otimes |\!\uparrow\rangle$,} and \smash{$|b\!\downarrow\rangle = |b\rangle \otimes |\!\downarrow\rangle$}. In the vicinity of the intersection, they are separated by an energy gap \smash{$\Delta=\varepsilon_{a \uparrow} - \varepsilon_{b \downarrow}$.} With details provided in Part II of the Supplemental Material \cite{supp}, we develop an analytic theory under the sole assumption that the other bands are distant in energy on the SOI scale and can be neglected. We show that the SOI-induced coupling between the two crossing bands  is encoded by a momentum-dependent real vector \smash{${\bm \beta}_{a b}$}, independent of the direction of the magnetization $\bm m$, and reads \smash{$|V_{\scriptscriptstyle SOI}| = |{\bm m} \times {\bm \beta}_{a b}|$.} This lifts the degeneracy, resulting in a pair of reconnected bands with energies \smash{$\varepsilon_{\pm} = (\varepsilon_{a \uparrow} + \varepsilon_{b \downarrow})/2 \pm \frac{1}{2} \sqrt{\Delta^2 + 4 |V_{\scriptscriptstyle SOI}|^2}$.} We define the dimensionless spin polarization of a Bloch state $|\alpha\rangle$ as \smash{$\bm s_\alpha = \langle \alpha | \hat{\bm \sigma} | \alpha \rangle$}, with  \smash{$\frac{\hbar}{2} \hat{\bm \sigma}$} the spin operator. It splits into a longitudinal part \smash{$s^{\parallel}_\alpha = {\bm m} \cdot {\bm s_\alpha}$} and a transverse part \smash{$\bm s^{\perp}_\alpha = {\bm m} \times ( \bm s_\alpha \times {\bm m} )$}. We recover the previously derived reversal of the longitudinal spin polarization  \cite{gradhand:2009} {\em i.e.}, \smash{$s^{\parallel}_{\pm} = \mp \Delta / \sqrt{\Delta^2 + 4 |V_{\scriptscriptstyle SOI}|^2}$.} Most importantly, we uncover a transverse spin polarization
\begin{equation}
    {\bm s}^{\perp}_{\pm} =  \mp \frac{2 \, | \langle a | b \rangle | \, \bm m \times \bm \beta_{ab}}{\sqrt{\Delta^2 + 4 | \bm m \times \bm \beta_{ab} |^2}} . \label{eq:s_perp}   
\end{equation}
Hence, {\em we predict the generic existence of an SOI-induced transverse spin texture at the seams of the spin-mosaic Fermi surfaces of centrosymmetric ferromagnetic metals}. By direct inspection of Eq.(\ref{eq:s_perp}), one can see that its maximum amplitude occurs on the seam center line, and is given by the orbital overlap $|\langle a | b \rangle|$ of the two bands, independently of the strength of the SOI. Consequently, it is possibly close to unity in some cases. Also, the lateral extension of the seams, which determines in part the number of states at the Fermi surface carrying a transverse spin polarization, increases with the SOI strength, or when the gap varies slowly in the vicinity of the seams. This latter condition calls for Fermi surfaces nearly tangent to each other close to their avoided crossing, which is more likely when the exchange splitting is small.

\begin{figure}[t]
\vspace*{10.3 pt}
\includegraphics[width=\columnwidth]{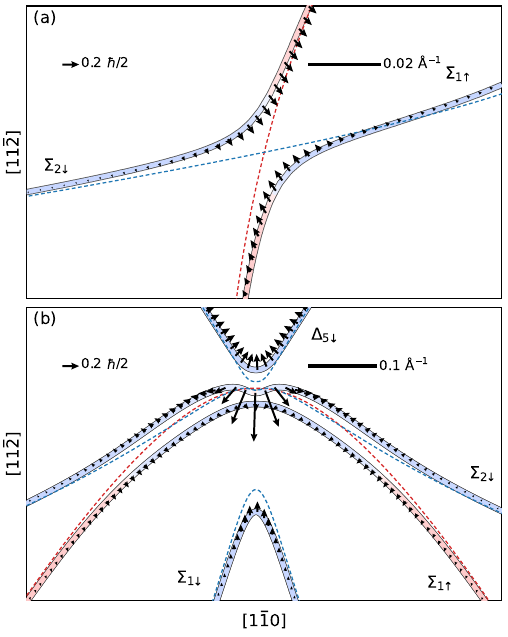}
\vspace{-6 pt}
\caption{\label{fig:seam} {\em Reconnection of the Fermi surfaces and transverse spin texture of fcc Ni,} in two cut-planes perpendicular to ${\bm m} \parallel [111]$: (a) the plane through $\Gamma$, at an isolated crossing, the panel centered at the wave vector $(-0.54, 0.98, -0.43)$~\AA$^{-1}$; (b) the plane at $-0.80$~\AA$^{-1}$ from $\Gamma$ along ${\bm m}$, through the point of largest transverse spin polarization in Ni, the panel centered at $(-0.04, -0.06, -1.29)$~\AA$^{-1}$. Dashed lines: the Fermi lines without SOI, majority in red and minority in blue. Full lines: the Fermi lines with SOI, with the tint of Fig.~\ref{fig:1} for the spin polarization along ${\bm m}$. Arrows: the transverse spin polarization.}
\vspace{-12 pt}
\end{figure}

These qualitative predictions, derived from a simple two-band theory,  are largely confirmed by {\em ab initio} computations for the centrosymmetric elemental ferromagnetic metals bcc Fe, hcp Co and fcc Ni. With further details provided in Part I of the Supplemental Material \cite{supp} (and Refs.~\cite{perdew:1996,hobbs:2000,ma:2015,souza:2001,marzari:1997,lorensen:1987,kittel:2004,daalderop:1990} therein), we have computed with VASP \cite{kresse:1993,kresse:1994,kresse:1996b,kresse:1996} the electronic structure at constrained magnetization directions, with and without SOI, followed by Wannier interpolation for both the Hamiltonian and the spin operator, using Wannier90 \cite{mostofi:2008,pizzi:2020}. This allows to map the full spin polarization on the Fermi surfaces. In Fig.~\ref{fig:1}, we show, for each metal, the spin mosaics carrying the strongest transverse spin texture. They nearly touch at their seams, close to the lines at which they intersect in the absence of SOI. The transverse spin polarization is confirmed to be concentrated on these seams, with a smoothly varying orientation, even in momentum, as imposed by symmetry. With $\bm m$ along an easy axis, it reaches large maximum amplitudes {\em i.e.}, $0.64$ in Fe, $0.56$ in Co and $0.55$ in Ni; all $O(1)$ values, even though the SOI opens only small gaps of $20$ to $60$~meV at the seams. When only two bands are involved, as in Fig.~\ref{fig:seam}(a), the transverse spin texture exhibits an antiparallel arrangement, as prescribed by Eq.(\ref{eq:s_perp}). The texture becomes more complex when additional bands are in close proximity, as shown in Fig.~\ref{fig:seam}(b), around one of the points of maximum transverse polarization in Ni. The seams, when defined (arbitrarily) by \smash{$|s^{\parallel}_\alpha| < 0.9$,} are found to cover $12\%$, $6\%$ and $17\%$ of the Fermi surface area in Fe, Co and Ni, respectively. Co being the exception, this confirms some correlation between a smaller exchange and wider seams, the SOI strength being comparable in the three metals while the average exchange splitting is reduced by a factor of four from Fe to Ni. 

{\em Current-gradient-induced spin-orbit torques.---}  In magnetic systems with broken inversion symmetry, spin textures odd in momentum are intimately associated with current-induced SOT \cite{manchon:2008,miron:2010,haney:2013,manchon:2015,manchon:2019}. Importantly, the transverse spin texture that we have uncovered in centrosymmetric ferromagnetic metals cannot be associated with a torque under a uniform current \cite{garate:2009}. However, this symmetry restriction does not apply to the response to an inhomogeneous current, since a current gradient is parity-even. Hence, {\em  we predict the existence of current-gradient-induced SOTs in centrosymmetric ferromagnetic metals}. This prediction matters because current density gradients are ubiquitous in any nanoscale conductor, for instance in thin films, within kinetic layers extending over a few electron mean free paths from any surfaces or interfaces \cite{fuchs:1938,sondheimer:1952,camley:1989}. This predicted current-gradient-induced SOT {\em within ferromagnetic conductors} is entirely distinct from recently hypothesized torques induced by the orbital Hall effect \cite{go:2020,hayashi:2023}, the interfacial orbital Edelstein effect \cite{krishnia:2023,nikolaev:2024}, the spin-vorticity coupling \cite{okano:2019,nakayama:2025,yi:2025} or the anomalous SOT \cite{wang:2019,amin:2020}. 

A quantitative microscopic theory follows from quantum kinetic theory \cite{valet:2023,valet:2025}. At the leading order in the semiclassical expansion, which corresponds to the intraband response, and with further justifications provided in the End Matter, a general expression for the current-gradient-induced torque reads
\begin{equation} \label{eq:tau_fund}
{\bm \tau}_{\scriptscriptstyle ib}^{\scriptscriptstyle E} = \bm m \times \sum_\alpha \int_{p} n'_{\scriptscriptstyle FD} (\varepsilon_\alpha) \, \phi_\alpha^{\scriptscriptstyle E} \, \langle \alpha | \hat{\mathcal J} \otimes \hat{\bm \sigma} | \alpha \rangle ,
\end{equation}
with \smash{$\hat{\mathcal J}$} the exchange operator, which can be computed from numerical {\em ab initio} Hamiltonians, as detailed in Part I of the Supplemental Material \cite{supp} (see also Ref.~\cite{saito:2024} therein). With further details provided in Part III of the Supplemental Material \cite{supp}, a more transparent connection with the transverse spin texture of Eq.~(\ref{eq:s_perp}) is then obtained at the level of approximation of the previously introduced two-band theory, and it reads
\begin{equation} \label{eq:tau_centro}
    {\bm \tau}_{\scriptscriptstyle ib}^{\scriptscriptstyle E} \approx - \sum_{\pm} \int_{p} n'_{\scriptscriptstyle FD}(\varepsilon_\pm) \, \phi_\pm^{\scriptscriptstyle E} \, \frac{ \bm m \cdot \bm \beta_{ab} }{ | \langle a | b \rangle | } \, \bm m \times ( \bm m \times \bm s_{\pm} ) .
\end{equation}
In both Eq.~(\ref{eq:tau_fund}) and Eq.~(\ref{eq:tau_centro}), \smash{$\int_{p} \equiv \int_{\scriptscriptstyle BZ} d^3 p / (2 \pi \hbar)^3$} is the momentum integral over the Brillouin zone ($BZ$), while  \smash{${f}^{\scriptscriptstyle{E}}_{\alpha} = - n'_{\scriptscriptstyle FD}(\varepsilon_\alpha) \, \phi_\alpha^{\scriptscriptstyle E}$} is the change in the distribution function in linear response to the electric field $\bm E$, in which \smash{$n_{\scriptscriptstyle FD}$} is the Fermi-Dirac distribution, and \smash{$\phi_\alpha^{\scriptscriptstyle E}$} the deviation function \cite{ziman:1960}. The formal similarity between the torque from Eq.~(\ref{eq:tau_centro}) and the Landau-Lifshitz damping torque \cite{landau:1935} is striking. As a matter of fact, since \smash{$\bm \beta_{ab}$} does not depend on \smash{$\bm m$} while \smash{$\bm s_{\pm}$} reverses with it, and since the deviation is independent of \smash{$\bm m$} at the leading order in the SOI, the torque of Eq.~(\ref{eq:tau_centro}) is even in \smash{$\bm m$}. Hence, it constitutes a dissipative torque in the classification of Ref.~\cite{hals:2013}, damping or antidamping since it is odd in the applied field. We also stress that, while Eq.~(\ref{eq:tau_fund}) seems to make the exchange the mediator of the torque exerted on the magnetization by the spin density in linear response to the field, Eq.~(\ref{eq:tau_centro}) establishes that it is in fact exerted through the SOI. And since \smash{$\bm m \cdot \bm \beta_{ab}$} and the seam width are both of first order in the SOI, \smash{${\bm \tau}_{\scriptscriptstyle ib}^{\scriptscriptstyle E}$} is at least of second order in the SOI strength. The same holds for the equilibrium torque {\em i.e.}, the magnetocrystalline anisotropy torque, which Eq.~(\ref{eq:tau_fund}) also yields if the occupation function is set as the equilibrium one. It is then of at least second order in the SOI, as previously recognized \cite{bruno:1989,vanderlaan:1998}. However, its second-order part vanishes by symmetry in cubic crystals. In a centrosymmetric ferromagnetic metal, the current-gradient-induced torque and the anisotropy torque are thus the same microscopic SOT acting on the magnetization, originating from the linear-in-field deviation and from the equilibrium occupation, respectively. The same connection was previously established for the SOT originating from a uniform current in systems with broken inversion symmetry in Ref.~\cite{garate:2009}.

Let us now clarify the connection between \smash{${\bm \tau}_{\scriptscriptstyle ib}^{\scriptscriptstyle E}$} and the charge current gradient. In both Eq.~(\ref{eq:tau_fund}) and Eq.~(\ref{eq:tau_centro}), the part of the integrand other than the deviation is even in momentum by inversion symmetry. Hence, the deviation must also have a part even in momentum for the torque to be nonzero. Critically, this is not the case in the bulk of a homogeneous conductor carrying a uniform current, where it is purely odd. Conversely, within the kinetic layer near boundaries, the deviation acquires a component even in momentum associated with the current gradient along the normal, realizing the predicted current-gradient-induced torque. The kinetic layer is captured by the linear Boltzmann equation \cite{ziman:1960}
\begin{equation} \label{eq:boltz_film}
    v^z_\alpha \, \partial_z \phi_\alpha^{\scriptscriptstyle E} + e\, \bm E \cdot \bm v_\alpha = \mathcal{C}_\alpha[{\bm \phi^{\scriptscriptstyle E}}] ,
\end{equation}
specialized here for a thin film whose normal is along \smash{$O_z$}, with \smash{$\bm v_\alpha = \partial_{\bm p}\varepsilon_\alpha$} the band velocity and \smash{$\mathcal{C}_\alpha$} the collision integral. As further detailed in Part IV of the Supplemental Material \cite{supp} (see also Refs.~\cite{kohn:1957,soffer:1967,cercignani:1988,penn:1999,saye:2015,gilat:1966,lehmann:1972,blochl:1994,atkinson:1997,saad:1986,saad:2003,morgan:1995,hopf:1934,chandrasekhar:1950} therein), we solve Eq.(\ref{eq:boltz_film}) with a collision integral corresponding to scalar impurity disorder. As for the boundary conditions, we allow in each band for the back scattered states to retain {\em on average} a fraction $0 \le p \le 1$ of the in-plane current carried by the incoming states, in close analogy with the partial accommodation boundary condition introduced by Maxwell for rarefied gases \cite{maxwell:1879}. This provides a way to interpolate continuously between two limits: a fully diffusive surface for $p=0$, and a surface conserving the in-plane current for $p=1$. Hence, it constitutes a natural generalization to multiband metals of the boundary conditions originally considered by Fuchs and Sondheimer \cite{fuchs:1938,sondheimer:1952}. Then, Eq.(\ref{eq:boltz_film}) can be efficiently solved numerically for arbitrary film thickness and crystallographic orientation. As the collision integral couples the states only through the moments of the deviation on the band projectors over the Fermi surface, Eq.~(\ref{eq:boltz_film}) reduces to a system of coupled differential equations in the $z$ direction for the moments. It is solved in two successive stages, the relaxation-time solution in closed form followed by the in-scattering contribution on a mesh. 

\begin{figure}[t]
\includegraphics[width=\columnwidth]{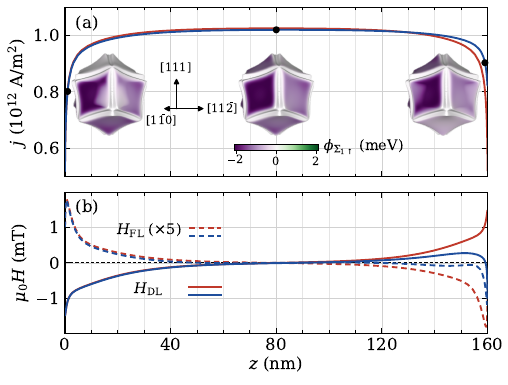}
\vspace{-20 pt}
\caption{\label{fig:current} {\em Current density and effective field,} in a 160~nm thick $(111)$ fcc Ni film with ${\bm m}$ and ${\bm E}$ along $[1\bar{1}0]$, for an average current density of $10^{12}$~A/m$^2$. Red lines: $p=0$ at bottom and top. Blue lines: $p=0$ at bottom and $p=0.5$ at top. (a) Current density profiles; and, for the asymmetric film, insets with the deviation $\phi_{\Sigma_{1\uparrow}}$ on the $\Sigma_{1\uparrow}$ Fermi surface, at 1~nm from the bottom, at the center and at 1~nm from the top (black dots). The triad of crystal axes defines the viewpoint. (b) The effective field components, damping-like, $H_{\rm DL}$ along $[111]$, and field-like, $H_{\rm FL} (\times 5)$ along $[11\bar{2}]$.}
\vspace{-8 pt}
\end{figure}%
In Fig.~\ref{fig:current}(a), we display the current density profiles obtained from the solution of Eq.(\ref{eq:boltz_film}) for a thick Ni film, with two fully decoupled kinetic layers. Their thickness, here of the order of 20~nm, is controlled by the bulk scattering strength, calibrated to a bulk resistivity of $12.4~\mu\Omega\,$cm in this case. The current density falls to $52\%$ of its average value at a surface with $p=0$, while it keeps $82\%$ of it at a surface with $p=0.5$. As shown in the insets, the deviation is distorted within the kinetic layers, to an extent controlled by $p$, acquiring a part even in momentum absent in the bulk. The depth profiles for the resulting damping-like (DL) and field-like (FL) components of the effective field, \smash{$\bm H = -(1/M_{\scriptscriptstyle S}) \bm m \times {\bm \tau}_{\scriptscriptstyle ib}^{\scriptscriptstyle E}$,} as computed from Eq.~(\ref{eq:tau_fund}), are shown as solid and dashed lines, respectively, in Fig.~\ref{fig:current}(b), with the DL dominating. When the ferromagnetic film has identical top and bottom surfaces, both the current density gradients and the torques are exactly antisymmetric. This is shown in Fig.~\ref{fig:current} with the red lines, obtained for $p=0$ at both surfaces. In such a case, the net total torque is zero. Evidently, a net torque requires an asymmetric film, for instance with $p=0.5$ at the top and $p=0$ at the bottom, resulting in the blue lines shown in Fig.~\ref{fig:current}. As an element of comparison for the relative strength of the effect, under the same conditions, we report an average DL field over the first 20 nm from the bottom surface of $-0.10$~mT in $(001)$ Fe, $-0.09$~mT in $(0001)$ Co, against the $-0.63$~mT in $(111)$ Ni that one can compute from the data of Fig.~\ref{fig:current}(b). This is positively correlated with the relative extensions of the seams over the Fermi surfaces reported in the previous section, as expected.

\begin{figure}[t]
\includegraphics[width=\columnwidth]{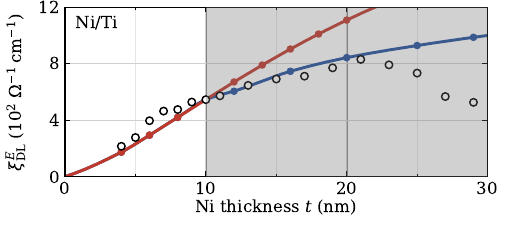}
\vspace{-20 pt}
\caption{\label{fig:hayashi} {\em Ni layer thickness dependence of the damping-like torque efficiency,} for a series of Ti(8~nm)/Ni($t$) bilayers measured in Ref.~\cite{hayashi:2023} (open circles, data extracted from Fig.~2d therein), and computed from our model of intraband current-gradient-induced torque (red and blue dots at the computed thicknesses, with the continuous lines as guides to the eye).}
\vspace{-8 pt}
\end{figure}

{\em Comparison with experiments.---} In thin-film heterostructures pairing a ferromagnetic metal layer (FM) with a low-SOI normal metal layer (NM), unexpectedly large current-induced torques have been measured, see Ref.~\cite{fukami:2026} for a review, and attributed to an OHE current in NM \cite{bernevig:2005,kontani:2009,go:2018b} exerting a torque on the magnetization once absorbed by FM \cite{go:2020}. While our work cannot rule out this interpretation, it is indisputable that in all these experiments current gradients exist near boundaries, while FM carries a transverse spin texture on its Fermi surface. Hence, a current-gradient-induced SOT, generated within FM, yet influenced by NM through the kinetic boundary conditions, may well be a significant contributor to the measured torques.

With the details reported in Part V of the Supplemental Material \cite{supp} (see also Refs.~\cite{swagten:1996,fukuzawa:2002} therein), we test this hypothesis against data from Ref.~\cite{hayashi:2023}, measured on a series of Ti/Ni bilayers {\em i.e.}, [substrate/\allowbreak Ti(8~nm)/\allowbreak Ni($t$)/\allowbreak SiO$_2$(4~nm)]. With the sign convention of Ref.~\cite{hayashi:2023}, the data we analyze are the DL torque efficiencies,  \smash{$\xi^{E}_{\scriptscriptstyle DL} = -(2e/\hbar)(M_{\scriptscriptstyle S}/E) \int_0^t H_{\scriptscriptstyle DL}\, dz$}. We treat the Ti layer as an effective diffusive bottom interface, setting $p=0$ there throughout, a justified hypothesis. With two constraints {\em i.e.}, the measured \smash{$\xi^{E}_{\scriptscriptstyle DL}$} and the resistivity for the $t=10$~nm sample estimated from measurements, and two fitting parameters {\em i.e.}, the top surface \smash{$p_s$} and the bulk resistivity \smash{$\rho_{Ni}$,} an exact match for $t=10$~nm is achieved, with the fitted values $p_s=0.80$ and $\rho_{Ni} = 12.4~\mu\Omega\,$cm. Without remaining fitting parameters, the DL efficiency is computed from this first model for a series of Ni thicknesses, resulting in the red dots and line in Fig.~\ref{fig:hayashi}. A quantitative agreement is achieved between the model and the data for $t \le 10$~nm, with rms relative differences of only 17\%. For $t \ge 10$~nm the measured efficiency saturates more quickly than this first model prediction. This can be reasonably explained for \smash{$10 \le t \le 20$}~nm by calibrating \smash{$p_s$} on a second series of data from \cite{hayashi:2023} for the samples [substrate/Ni($t$)/SiO$_2$(4~nm)], with the fitted $p_s(t)$ at the top boundary decreasing to $0.60$ at $t = 20$~nm, resulting in the blue dots and line from this adjusted model in Fig.~\ref{fig:hayashi}. For $t > 20$~nm, further structural changes of the samples probably cause the rapid drop of the efficiency.     

{\em Conclusions and perspectives.---} We have established from first principles that a prominent spin texture, transverse to the magnetization, generically exists on the spin mosaic Fermi surfaces of centrosymmetric ferromagnetic metals. Its maximum modulus is independent of the SOI. However, it is important to emphasize that the number of states exhibiting a significant transverse spin polarization increases with increasing SOI or decreasing exchange splitting. These predictions have been confirmed with {\em ab initio} computations for the elemental ferromagnets bcc Fe, hcp Co and fcc Ni, as illustrated in Figs.~\ref{fig:1} and \ref{fig:seam}. A current-gradient-induced SOT, which we have uncovered here for the first time, is associated with this transverse spin texture. This new SOT was shown, through a detailed analysis, to account for the DL torque measured in Ref.~\cite{hayashi:2023} on Ti/Ni bilayers. This finding questions the relevance of the OHE as an explanatory mechanism for the torques observed in NM/FM heterostructures with low-SOI NM layers.
\newline\hspace*{1em} Beyond this, one expects the current-gradient-induced SOT to be present in essentially all heterostructures with nanoscale ferromagnetic conductors, notably those with engineered resistivity gradients, in which significant current-induced torques have been recently measured   \cite{nakayama:2025,yi:2025}, and interpreted so far as evidence of spin-vorticity coupling \cite{matsuo:2013,takahashi:2016,kobayashi:2017}. 
\newline\hspace*{1em} We also want to stress that the spin mosaic picture of reconnected Fermi surfaces with opposite spin character, and the existence of SOI-induced transverse spin textures in inversion-symmetric systems, are not restricted to the ferromagnetic order. For instance,  it can be shown that these concepts carry over to metallic altermagnetic systems (see End Matter). Indeed, it appears that computational evidence of a spin texture transverse to the N\'eel vector at the seams of the Fermi surface of the metallic altermagnet CrSb has already been found (Supplementary Fig.~12 of Ref.~\cite{yang:2025} and its caption). While the implications were not discussed by the authors, our work suggests that this actually opens the possibility of a current-gradient-induced torque for the electrical control of the N\'eel  vector in altermagnets. 
\newline\hspace*{1em} In conclusion, we would like to emphasize that our findings open a new era for the development of current-gradient-induced SOT based devices, with a potentially major impact for the future of non-volatile electronics. Realizing this potential in full will require the combined engineering of (i) the newly uncovered Fermi surface spin textures of centrosymmetric magnetically ordered metals, through the optimization of the crystal structure and composition, and of (ii) the electron flow in finite geometries, through the manipulation of interface properties, compositional gradients or shape at the nanoscale. The development of novel real-space quantum transport simulation frameworks is likely to be key for accelerating this type of complex multiscale optimizations.    

\section{End Matter}
On the matter of the general expression of the torque density in linear response to an electric field, we recall that when neglecting Gilbert damping, $\bm m$ obeys the Landau-Lifshitz equation \smash{$(M_{\scriptscriptstyle S}/\gamma)\, \partial_t \bm m = \bm \tau$}, with \smash{$\bm \tau \equiv - \bm m \times ( \delta \mathcal{F} / \delta \bm m )$} the torque density, $\mathcal{F}$ the thermodynamic potential, $M_{\scriptscriptstyle S}$ the saturation magnetization and $\gamma (<0)$ the gyromagnetic ratio \cite{brown:1963,andreev:1980}. For a uniform $\bm m$ and in equilibrium, the Hellmann-Feynman theorem \cite{hellmann:1933,feynman:1939,pons:2020} then yields $\delta\mathcal F/\delta\bm m = \langle\partial_{\bm m}\hat h\rangle$, with \smash{$\hat h$} the Hamiltonian and the brackets indicating a thermal average, identifying the torque operator \smash{$\hat{\bm\tau} = -\bm m\times\partial_{\bm m}\hat h$} \cite{garate:2009}. This operator being the generalized force conjugate to $\bm m$, its expectation value still defines the torque beyond equilibrium, for its adiabatic part \cite{berry:1993,bode:2011}, leading to
\begin{equation}
{\bm \tau}^{\scriptscriptstyle E} = - \bm m \times \! \int_{p} {\rm Tr} \left[ \hat{\rho}^{\scriptscriptstyle{E}} \, \partial_{\bm m} \hat{h} \right] . \label{eq:torque_gen}
\end{equation}
As detailed in Part I of the Supplemental Material, for the elemental ferromagnets Fe, Co and Ni, \smash{$\hat{h}$,} as computed from first principles, can be accurately cast in a form whose only dependence on $\bm m$ is a rigid-exchange term, so that we can write \smash{$\partial_{\bm m} \hat{h}= \hat{\mathcal J} \otimes \hat{\bm \sigma}$}, with \smash{$\hat{\mathcal J}$} the exchange operator. At the leading order in the semiclassical expansion for \smash{$\hat{\rho}^{\scriptscriptstyle{E}}$,} {\em i.e.}, at order \smash{$\hbar^0$,} this yields Eq.~(\ref{eq:tau_fund}) of the main text, for the current-gradient-induced torque. There are of course quantum corrections to this torque. At order \smash{$\hbar^1$,} they can be obtained by substituting the coherences \smash{$\hat{\rho}^{\scriptscriptstyle{E}}_{\alpha \beta}$,} as derived in Ref.~\cite{valet:2025}, into Eq.~(\ref{eq:torque_gen}), yielding three distinct contributions. The first one is a Fermi sea torque \smash{$
{\bm \tau}_{\scriptscriptstyle in}^{\scriptscriptstyle E} = 2 \, (e E_i) \, \bm m \times \sum_{\substack{\alpha \\ \beta \neq \alpha}} \int_{p} n_{\scriptscriptstyle FD}(\varepsilon_\alpha) \, \bm{\Omega}^{i}_{\alpha \beta}$,}
controlled by the interband mixed Berry curvature
\begin{equation} \label{eq:Tgeom}
    {\bm \Omega}^{i}_{\alpha \beta} = - \hbar \, \frac{ {\rm Im} \big[ \langle \alpha | \partial_{p^i} \hat{h} | \beta \rangle \langle \beta | \partial_{\bm m} \hat{h} | \alpha \rangle \big] }{ ( \varepsilon_\alpha - \varepsilon_\beta )^2 } .
\end{equation}
It is the well-known intrinsic spin-orbit torque \cite{kurebayashi:2014,freimuth:2014,hanke:2017}, vanishing under inversion. As expected, the first non-vanishing contribution is then a quantum-gradient contribution, sourced by the spatial gradients of the deviation. Its existence was first predicted in Ref.~\cite{valet:2025b}, and we have now derived its explicit expression 
\begin{equation} \label{eq:Tgrad}
    {\bm \tau}_{\scriptscriptstyle qg}^{\scriptscriptstyle E} = \bm m \times \sum_{\substack{\alpha \\ \beta \neq \alpha}} \int_{p} n'_{\scriptscriptstyle FD}(\varepsilon_\alpha) \, \partial_z \phi_\alpha \, ( \varepsilon_\alpha - \varepsilon_\beta ) \,{\bm \Omega}^{z}_{\alpha \beta} ,
\end{equation}
involving the gap-weighted mixed Berry curvature. However, \smash{$\partial_z \phi_\alpha$}, as derived from the solution of Eq.~(\ref{eq:boltz_film}), becomes singular in the grazing velocity limit. This is a non-physical divergence due to the neglect of the quantization of the momentum in the perpendicular direction, and this calls for a careful regularization. A third contribution corresponds to the disorder vertex correction in the ladder approximation. A more detailed discussion of these quantum corrections to the current-gradient-induced torque is left to future work.

On the matter of the relevance of our findinds for altermagnets, we shall recall that, if one neglects the SOI, the Hamiltonian of an altermagnet commutes with \smash{$\bm n \cdot \hat{\bm \sigma}$,} with $\bm n$ the unit vector along the N\'eel vector. Hence, every band is of pure spin character, and two bands of opposite spin are generically degenerate on two-dimensional surfaces in the BZ, as in the ferromagnetic case. The two spin Fermi surfaces, images of each other under the crystal rotation exchanging the two sublattices, intersect along lines. The SOI couples the crossing pairs of bands with an orbital overlap $\langle a | b \rangle$ which is generically nonzero, since the exchange field is opposite on the two sublattices. One specificity of the altermagnetic case is that part of the crossings is enforced by symmetry: part of the intersection of the opposite-spin Fermi surfaces lies on the spin-degenerate nodal planes of the spin group. As in the ferromagnetic case, the SOI lifts the intersections everywhere but at isolated points, into pairs of states of vanishing longitudinal and large transverse spin polarization. This is visible in the published band structure of CrSb with SOI on the nodal plane $k_z = 0$ (Supplementary Fig.~12 of Ref.~\cite{yang:2025}): the spin polarization along the N\'eel vector vanishes by symmetry on every band, while the in-plane components reach their maximal value, $\pm 1/2$, on bands within $0.5$~eV of the Fermi level. In a centrosymmetric altermagnet, inversion maps each sublattice onto itself, as in CrSb \cite{li:2025}, and the transverse texture is even in momentum. Hence, a uniform current induces no torque on the N\'eel vector by symmetry, and the current-gradient-induced SOT becomes the leading candidate for its electrical control.

\begin{acknowledgments}
All authors thank Albert Fert for fruitful discussions. T.V. additionally thanks Gerrit Bauer, Shunsuke Fukami, Yukio Nozaki, and Sadamichi Maekawa for valuable discussions. This research has been supported by the France 2030 government grant managed by the French National Research Agency PEPR SPIN Grant No. ANR-22-EXSP0009 (SPINTHEORY), by the IFCPAR/CEFIPRA grant No. AWC15615 (RP04192F) and by the EIC Pathfinder OPEN Grant No. 101129641 ``OBELIX''. The {\em ab initio} calculations were performed using computational resources provided by GENCI–IDRIS under grant No. 2026-A0190912036. The use of generative artificial intelligence tools in this work is disclosed in Part VI of the Supplemental Material \cite{supp}.
\end{acknowledgments}

\bibliography{berry,berry2}

\end{document}